\documentclass{llncs}
\usepackage{hyperref}
\usepackage{makeidx}  
\usepackage{hhline} 
\usepackage[table]{xcolor} 
\usepackage[nomessages]{fp} 
\usepackage{amsmath}
\usepackage{amssymb}
\usepackage{mathtools}
\usepackage{xifthen}
\usepackage{ifthen}
\usepackage{tabularx}
\newcolumntype{C}{>{\centering\arraybackslash}X} 

\usepackage{pdflscape}
\usepackage{afterpage}
\usepackage{placeins}
\usepackage{dsfont}

\newif\iflong
\longfalse

\newif\ifcolor
\colortrue

\usepackage[belowskip=0pt,aboveskip=0pt,center]{caption}
\ifcolor
\newcommand{\cc}[1]{
\ifthenelse{\lengthtest{#1pt > 0pt}}{\FPeval{\red}{1}\FPeval{\green}{clip((1-(#1)))}\FPeval{\blue}{clip((1-(#1)))}}{\FPeval{\red}{clip((1+(#1))*0.8+0.2)}\FPeval{\green}{clip((1+(#1))*0.8+0.2)}\FPeval{\blue}{1}}\xdef\red{\red}\xdef\green{\green}\xdef\blue{\blue}\cellcolor[rgb]{\red, \green, \blue}#1}

\else
\newcommand{\cc}[1]{\FPeval{\red}{clip(((#1)+1)*0.8/2+0.2)}\FPeval{\green}{clip(((#1)+1)*0.8/2+0.2)}\FPeval{\blue}{clip(((#1)+1)*0.8/2+0.2)}\xdef\red{\red}\xdef\green{\green}\xdef\blue{\blue}\cellcolor[rgb]{\red, \green, \blue}#1}

\fi

\usepackage{wasysym}

\newcommand{\overlap}{$\ocircle$\hspace{-0.16cm}$\ocircle$\hspace{0.1cm}}

\title{On the evaluation potential of quality functions in community detection for different contexts}
\titlerunning{Comparison of quality functions}  
\author{Jean Creusefond\inst{1} \and Thomas Largillier\inst{1} \and Sylvain Peyronnet\inst{2}}
\institute{Normandy University,
\and
Qwant \& ix-labs}

\begin{document}
\maketitle              

\begin{abstract}
Due to nowadays networks' sizes, the evaluation of a community detection
algorithm can only be done using quality functions. These functions measure
different networks/graphs structural properties, each of them corresponding
to a different definition of a community.  Since there exists many
definitions for a community, choosing a quality function may be a
difficult task, even if the networks' statistics/origins can give some clues
about which one to choose.

In this paper, we apply a general methodology to identify different
\texttt{contexts}, \textit{i.e.} groups of graphs where the quality functions behave 
similarly. In these \texttt{contexts} we identify the best quality functions,
\textit{i.e.}  quality functions whose results are consistent with
expectations from real life applications.  

\end{abstract}

\section{Introduction}
Every community detection algorithm is justified by the search for
particular substructures, \textit{i.e.} communities, defined by a
particular purpose in a particular network. 
This combination of structured data and purpose makes the field
complex and fuzzy, but drives research to unravel the different
meanings that the word "community" bears.

As a result, a large number of desirable properties of communities
have been discovered.
To measure them, many works aimed at designing functions quantifying
these properties in order to evaluate the goodness of a community. 
Called quality functions, these mathematical tools are not only useful
for evaluation purposes but can also be used in greedy algorithms as
community detection methods directly.

However, evaluating an algorithm may be difficult because it implies
choosing between quality functions that often output contradictory
results. The structural properties of the network and of the
communities being looked for may strongly differ from one case to the
other. It is then of the utmost importance to identify the right
quality function for each graph. In order to do that we define the
notion of \texttt{context} which is a group of graphs where quality
functions behave similarly. One then only needs to identify the right
quality function for a \texttt{context} and the means to identify
which \texttt{context} a graph is part of.

In this paper we identify some \texttt{contexts} for community
detection, and select quality functions that feature behavior that is
coherent with real-world data. 
To achieve this goal, we compare 10 functions from relatively recent
literature, using 10 datasets featuring ground-truth, 7
community detection algorithms and 2 extrinsic evaluation functions. 
We look at the correlation between quality functions and real-world
data : do they rank higher clusterings that are close to the
ground-truth, and conversely? 
We then identify \texttt{contexts} when quality functions rank different
graphs in the same way.

\section{Related work}
The rise of community detection as a research field has inevitably
given birth to a variety of works on meta analysis. 
They feature a wide range of methods, but all of them are aimed to
identify quality functions with desirable properties. 

Van Laarhoven and Marchiori~\cite{van_laarhoven_axioms_2014} designed
six axioms that qualify intuitive good behavior of quality functions. 
They show that modularity does not satisfy two of them, partly because
of the resolution limit~\cite{fortunato_resolution_2007}. 

Yang and Leskovec~\cite{yang_defining_2012} studied 12 quality
functions that could be applied at cluster level. 
They classified them into four groups depending on how they were
correlated when applied to real-world graphs, and these groups
corresponded to the measured structural property. 
They designed ``goodness metrics'' that measure only one property of a
cluster and compared how the quality functions fared in order to
identify what property were measured by which function.

Almeida \textit{et al.} ~\cite{almeida_is_2011} compared the result of
5 quality functions when applied to 5 real-world graphs. 
They applied 4 parameterized algorithms on these networks and changed
the parameters to get different number of communities.
They observed that some metrics have the tendency to favor bigger
clusters while others favor the opposite.

Our approach differs from previous works by the scale and purpose of
our work : to the best of our knowledge, we are the first to focus on
the identification of \texttt{contexts} for quality functions.
Chakraborty \textit{et al.}~\cite{chakraborty_permanence_2014} have 
already applied part of this methodology in the context of community
detection in order to experimentally demonstrate the efficiency of the
quality function they proposed.

\newpage

\section{Quality functions}
\label{sec:qualities}
Throughout the rest of this paper we use the following notations.

\begin{tabular}[center]{l@{\hskip 1mm}l}
  General & \\
  \hline
  $G = (V,E)$ & Undirected graph (set of vertices, edges) \\
  $n$, $m$ & \# of vertices ($=$ nodes) and edges \\
  $N_{v\in V}$ & Set of neighbors of a node $v$ \\
  $k_{v\in V}$ & \# of neighbors (degree) of a node $v$ \\
  $k_m$ & median degree\\
  \rule{0pt}{4ex}Set-specific & \\
  \hline
  $m(S\subseteq V)$ & \# of internal edges of $S$ \\
  $m(S\subseteq V, S'\subseteq V)$ & \# of edges with one end in $S$ and another in $S'$ \\
  $N_{v\in V, S\subseteq V}$ & Internal neighborhood of $v$ in $S$ \\
  $k_{S\subseteq V}$ & Size of a cluster $S$ \\
  $Vol(S\subseteq V)$ & Volume of a cluster $S$ (sum of the degrees of the vertices) \\
  $diam(S\subseteq V)$ & Internal diameter of a cluster $S$\\
  \rule{0pt}{4ex}Clusterings & \\
  \hline
  $\mathcal{C}, \mathcal{L}$ & Clustering, set of sets of nodes whose union is $V$ \\
  $C(v\in V), L(v\in V)$ & Set of clusters in which a node $v$ belongs to in $\mathcal{C}$/$\mathcal{L}$
\end{tabular}
\newline

A quality function is an application
$f(G, \mathcal{C}) \rightarrow \mathds{R}$, whose purpose is to
quantify the quality of a clustering on a graph.  
For brevity, we omit the graph input.
Note that quality functions are different from comparison methods,
the latter comparing two clusterings. 

In order to ease comparisons, we normalize some quality functions. 
We categorize the functions depending on the locality of information they use.
We identify three classes of locality : vertex-level, community-level
and graph-level. The formula for each quality function can be found in Tab.~\ref{tab:quality}.

\paragraph{Vertex-level quality functions} compute a quality for every
node in the graph and output the
average as the total quality of the clustering on the
graph. Let $v\in V$ be the considered node, and $C\in \mathcal{C}$ be
the community of $v$.

The \textbf{Local internal clustering
coefficient}~\cite{watts_collective_1998} (called clustering
coefficient from now on) of a node is the probability that two of his neighbors
that are in the same community are also neighbors. 
The clustering property of communities is actually one of the most
well-known in the field, and is explained by the construction of
social networks by homophily.

This property is included
in \textbf{Permanence}~\cite{chakraborty_permanence_2014}, where it is 
combined with a notion of equilibrium for the nodes concerning their
membership to their community.
A node has a lower Permanence if there is another community than its
own that highly attracts it, \textit{i. e.} to which it is very
connected compared to its connection to its community.

The \textbf{Flake-ODF}~\cite{flake_efficient_2000} compares internal
to external degree.
It is similar to the \textbf{Fraction Over Median
Degree}~\cite{yang_defining_2012} (FOMD), that compares internal
degree and the median degree in the whole graph.

\paragraph{Community-level quality functions} compute a score for each
cluster and output the sum as the quality of the clustering.

The \textbf{Conductance}~\cite{kannan_clusterings_2004} and
the \textbf{Cut-ratio} are concerned with the external connectivity of
the community.
The Cut-ratio normalizes it with the number of potential edges between
the individuals of the community and the remainder of the network.
On the other hand, the Conductance is normalized by the same potential
number of edges but takes into account the degrees in the community
(few edges may reach a community of consisting of a few nodes with
small degree).
We weight these local measures with the size of the community so that
each vertex in the networks has the same level of participation in the
measure.

The \textbf{Compactness}~\cite{creusefond_finding_2015} measures the
potential speed of a diffusion process in a community.
Starting from the most eccentric node, the function captures the
number of edges reached per time step by a perfect transmission of
information.
The underlying model defines community as a group of people within
which communication quickly reaches everyone.

\textbf{Modularity}~\cite{newman_finding_2004} is the difference between the number of internal edges of the community versus the expected number of edges.
This expectancy is expressed using the configuration model, a graph model guaranteeing the same degree distribution as the original one but with randomized edges.
Assuming that this model ignores community structure, a high difference between expectancy and reality would indicate an abnormal density, \textit{ergo} community structure.

\paragraph{Graph-level quality functions} output the score of the whole graph.
\textbf{Surprise}~\cite{aldecoa_surprise_2013} (in its asymptotical approximation~\cite{traag_detecting_2015}) and \textbf{Significance}~\cite{traag_significant_2013} are based on the computation of an asymmetric difference, the Kullback-Leibler divergence, between two-points probability distributions (only one event and its complement).
Considering $x$/$y$ indifferently as one of the two probabilities featured in the reference/non-reference distribution, the divergence is : $D(x||y) = x \log(x/y) + (1-x) \log((1-x)/(1-y))$.

The reference distribution of Surprise models the probability that an
edge is internal to a community, and the non-reference is the
event that a couple of nodes are inside the same community.
Significance features one reference distribution per community that
corresponds to the event that a random couple of nodes inside the same
community are linked by an edge.
The non-reference distribution is the same value for the whole graph.
\begin{table}[h]
  \caption{Quality functions\label{tab:quality}}
  \begin{tabularx}{\textwidth}{X|c}
    Name & Function \\
    \hline
    Local clustering coefficient & $f_{clus}(v, C) = \dfrac{2*|\{u\in N_{v, C}, w\in N_{v, C}\backslash\{u\}, (u,w)\in E\}|}{|N_{v, C}|(|N_{v, C}|-1)}$ \\
    \hline
    Permanence & $f_{perm}(v, C) = \dfrac{m(v, C)}{max_{C'\in \mathcal{C}\backslash \{C\}}(m(v, C'))\times k_v} + f_{clus}(v, C) - 1$ \\
    \hline
    1-Flake-ODF & $f_{flak}(v, C) = \begin{dcases*}
      1  & when $m(v, C) > m(v, V\backslash C)$\\
      0 & otherwise
    \end{dcases*}$ \\
    \hline
    FOMD & $f_{FOMD}(v, C) = \begin{dcases*}
      1  & when $m(v, C) > d_m$\\
      0 & otherwise
    \end{dcases*}$ \\
    \hline
    1-Cut ratio & ${f_{cut}(C) = \left(1 - \dfrac{m(c, V\backslash C)}{k_C(n - k_C)}\right)\times \dfrac{k_C}{n}}$ \\
    \hline
    1-Conductance & $f_{cond}(C) = \left(1 - \dfrac{m(v, V\backslash C)}{Vol(C)}\right)\times \dfrac{k_C}{n}$ \\
    \hline
    Compactness & $f_{comp}(C) = \dfrac{m(C)}{diam(C)}$ \\
    \hline
    Modularity & $f_{mod}(C) = \dfrac{m(C)}{m} - \left(\dfrac{Vol(C)}{2m}\right)^2$ \\
    \hline
    Surprise & $f_{surp}(\mathcal{C}) = D\left.\left.\left(\dfrac{\sum_{C\in \mathcal{C}}m(C)}{m} \right|\right| \dfrac{\sum_{C\in \mathcal{C}}\binom{k_C}{2}}{\binom{n}{2}} \right)$ \\
    \hline
    Significance & $f_{sign}(\mathcal{C}) = \sum_{C\in \mathcal{C}} \binom{k_C}{2} D\left.\left.\left(\dfrac{m(C)}{\binom{k_C}{2}} \right|\right| \dfrac{m}{\binom{n}{2}}\right)$
  \end{tabularx}
\end{table}

\section{Networks with ground-truth}
\label{sec:networks}
To identify \texttt{contexts} for community detection, we need some real-life
information on what a community actually is. 
We therefore pulled 10 networks with known community structure from
literature.
To compare them with the algorithms that classify all nodes, vertices
with no ground-truth communities are removed and only the largest
connected component is considered.  We note $\rightarrow$ for directed
networks and \overlap for overlapping communities.

\paragraph{Collaboration networks}(cf Tab.~\ref{tab:networks:collab}).
The networks represent people working together in certain
organizations. They have a strong underlying bipartite structure.
\footnotetext[1]{\url{http://snap.stanford.edu/data/}}
\footnotetext[2]{\url{http://konect.uni-koblenz.de}}
\footnotetext[3]{\url{https://github.com/blog/466-the-2009-github-contest}}
\noindent
\begin{table}
  \caption{Collaboration networks\label{tab:networks:collab}}
  \begin{tabularx}{\textwidth}{X|c|c|c|c|c|X}
    Name & $n$ & $m$ & nodes & edges & communities \\
    \hline
    DBLP\protect\footnotemark[1] \cite{yang_defining_2012} & 129981 & 332595 & authors & co-authorships & publication venues \overlap \\
    \hline
    CS \cite{chakraborty_computer_2013}\cite{chakraborty_citation_2014} & 400657 & 1428030 & authors & co-authorships & publication domains \overlap \\
    \hline
    Actors (imdb)\protect\footnotemark[2]~\cite{barabasi_emergence_1999} & 124414 & 20489642 & actors & co-appearances & movies \overlap \\
    \hline
    Github\protect\footnotemark[2]\protect\footnotemark[3] & 39845 & 22277795 & developers & co-contributions & projects \overlap
  \end{tabularx}
\end{table}

The Computer Science (CS) network comes from the same source as the DBLP network, but it features only computer scientists and a different kind of ground-truth.
Furthermore, the actors and github networks are constructed from bipartite graphs, and therefore form cliques inside of the communities.

\paragraph{Online Social Networks (OSNs)} (cf Tab.~\ref{tab:networks:osn}).
Most of these networks are originally directed but due to the high
reciprocity the original authors considered safe to set all links as
undirected.
\noindent
\begin{table}
  \caption{Online Social Networks\label{tab:networks:osn}}
  \begin{tabularx}{\textwidth}{c|c|c|c|c|c|X}
    Name & $n$ & $m$ & nodes & edges & communities \\
    \hline
    LiveJournal\protect\footnotemark[1] \cite{yang_defining_2012} & 1143395 & 16880773 & bloggers & following $\rightarrow$ & explicit groups \overlap \\
    \hline
    Youtube\protect\footnotemark[1] \cite{yang_defining_2012} & 51204 & 317393 & youtubers & following $\rightarrow$ & explicit groups \overlap \\
    \hline
    Flickr \cite{mislove_measurement_2007} & 368285 & 11915549 & users & following $\rightarrow$ & explicit groups \overlap
  \end{tabularx}
\end{table}


\paragraph{Social-related networks} (cf Tab.~\ref{tab:networks:social})
Nodes in these networks do not represent people, but their connections
are created by social interaction.
\noindent
\begin{table}
  \caption{Social-related networks\label{tab:networks:social}}
  \begin{tabularx}{\textwidth}{c|c|c|c|c|c|X}
    Name & $n$ & $m$ & nodes & edges & communities \\
    \hline
    Amazon\protect\footnotemark[1] \cite{yang_defining_2012} & 147510 & 267135 & products & frequent co-purchases & categories \\
    \hline
    Football \cite{girvan_community_2002} & 115 & 613 & football teams & $> 1$ one disputed match & divisions \\
    \hline
    Cora\protect\footnotemark[2] \cite{vsubelj_model_2013} & 23165 & 89.156 & scientific papers & citations $\rightarrow$ & categories
  \end{tabularx}
\end{table}



\paragraph{Artificial benchmarks} We use the
    Lancichinetti-Fortunato-Radicchi
    (LFR)~\cite{lancichinetti_benchmark_2008} benchmark as a
    validation method for our methodology.
    
On this benchmark, we may chose the number of nodes, the average
degree ($\hat{k}$), the maximum degree ($k_{max}$), the mixing
parameter ($\mu$), the coefficients of the power laws of degree and
community size distributions (respectively $t_1$ and $t_2$), the
average clustering coefficient ($\hat{cc}$), the number of nodes
belonging to multiple communities ($on$) and the number of communities
they belong to ($om$).
\noindent
\begin{table}
\centering
\caption{The parameters of the five classes of synthetic LFR networks}
\begin{tabularx}{0.95\textwidth}{C|c|C|C|C|C|C|C|C|C}
name & $n$ & $\hat{k}$ & $k_{max}$ & $\mu$ & $t_1$ & $t_2$ & $\hat{cc}$ & $on$ & $om$ \\
\hline
LFRa & 10 000 & 50 & 1 000 & 0.1 & 2.5 & 2.5 & 0.2 & 8 000 & 4 \\
\hline
LFRb & 100 000 & 50 & 2 500 & 0.1 & 2.5 & 2.5 & 0.2 & 8 000 & 4 \\
\hline
LFRc & 10 000 & 100 & 500 & 0.4 & 2.1 & 2.0 & 0.1 & 8 000 & 5 \\
\hline
LFRd & 10 000 & 50 & 1 000 & 0.1 & 2.5 & 2.5 & 0.2 & 0 & 0 \\
\hline
LFRe & 10 000 & 100 & 500 & 0.4 & 2.1 & 2.0 & 0.1 & 0 & 0 \\
\end{tabularx}
\label{table:LFR-parameters}
\end{table}

As presented in Tab.~\ref{table:LFR-parameters}, we have 5 classes of
networks. The \textit{a} class represents a standard social network,
with common values for each parameter. We note that the mixing
parameter is quite low (the communities should be well-cut) and the
communities are overlapping. 
The \textit{b} class is the same as the \textit{a} class but with ten
times more nodes. 
The \textit{c} class is however completely different, with all its
parameters changed except size (but it is still overlapping). 
The \textit{d}(\textit{e}) class is the same as
the \textit{a}(\textit{c}) class but without any overlapping
community. 

\section{Comparison methods}
\label{sec:comparison}
A comparison method (or extrinsic clustering evaluation
metric~\cite{amigo_comparison_2009}) is an application
$f(\mathcal{C}, \mathcal{L}) \rightarrow [-1;1]$, whose purpose is to
evaluate the closeness of two clusterings.

The \textbf{Normalized Mutual Information (NMI)} measures the quantity
of information gained by the knowledge of one clustering compared to
the other.
The version that we use was introduced by Lancichinetti \textit{et
al.}~\cite{lancichinetti_detecting_2009}. 

The \textbf{F-BCubed (fb3)}~\cite{bagga_entity_1998} precision
measures for each element $e$ the proportion of its associates
(\textit{e.g.} individuals that are in the same cluster) in
$\mathcal{C}$ that are still its associates in $\mathcal{L}$, and takes
the average among all $e$. 
Amig\'o \textit{et al.}~\cite{amigo_comparison_2009} extended this
metric for overlapping clustering, taking into account the number of
clusters in common that $e$ and its associates have.
They define BCubed overlapping precision and recall as follows :
\begin{eqnarray}
  &prec(C, L) = Avg_e\left[Avg_{\substack{e'\\C(e)\cap
  C(e') \ne \emptyset}}\left(\frac{min(|C(e)\cap C(e')|,|L(e)\cap
  L(e')|)}{|C(e)\cap C(e')|}\right)\right] \\
  &recall(C,L) = prec(L,C) \\
  &F\text{-}BCubed(C, L) = \dfrac{1}{\dfrac{1}{2*prec(C,L)}
  + \dfrac{1}{2*recall(C,L)}} 
\end{eqnarray}

Amig\'o \textit{et al.}~\cite{amigo_comparison_2009} also gave an
extensive comparison of evaluation metrics by designing intuitive
properties of goodness. 
Their conclusion was that the F-BCubed measure satisfied all of them,
while the other common metrics fail at least on of these axioms.

\section{Experimental setup}

In this section, we describe our experiments. We first cover the
methodology, then present the community detection algorithms used to
generate clusterings and finally tools we used to keep tractable the
number of operations.

\subsection{Methodology}
\label{sec:methodology}
The methodology has two goals : to identify \texttt{contexts} in which
quality functions behave in the same way, and to identify the best
quality functions for each \texttt{context}. 
For each graph with ground-truth communities
(cf Sec.~\ref{sec:networks}), we execute the following steps : 
\begin{enumerate}
\item \label{item:algs-apply} Apply various community detections methods on the base graph
(cf Sec.~\ref{sec:algorithms}) 
\item Compute quality functions over the resulting clusterings (cf
Sec.~\ref{sec:qualities}) 
\item Compare the communities found to the ground-truth, creating a gold
standard value for each clustering (cf Sec.~\ref{sec:comparison})
\item \label{item:spear-gold} Compare for each graph the ranking of the clusterings given by the gold standard value
to the ranking of clusterings measured by quality functions with Spearman's coefficient. For each graph, each quality now have a score.
\item \label{item:spear-qual} For each couple of graphs, compute the correlation of the previous scores using Spearman's coefficient.
\end{enumerate}

The rationale behind step~\ref{item:spear-gold} is that a quality
function fits a ground-truth if the clusterings that are the closest
to the ground-truth are highly ranked with the quality, and
conversely. 
Therefore, at this step we can conclude which quality function is
the best for each graph. 
We also need to go through step~\ref{item:spear-qual} in order to
identify \texttt{contexts}: the graphs are compared on their ranking from the
quality functions, and \texttt{contexts} may be identified as sets of graphs
that are highly correlated.

\subsection{Community detection algorithms}
\label{sec:algorithms}
Since we consider large graphs, we decided to use community detection
algorithms that have sub quadratic time and space complexity. We chose
several methods, based on their availability, efficiency, originality
and/or spread.

\iflong
\else

We classify the algorithms we use in three groups :
\begin{itemize}
\item Modularity optimization : Louvain~\cite{blondel_fast_2008},
Clauset~\cite{clauset_finding_2004}
\item Random walks : MCL~\cite{van_dongen_graph_2000},
Infomap~\cite{rosvall_maps_2008},\item Heuristics :
LexDFS~\cite{creusefond_finding_2015},
3-core~\cite{seidman_network_1983}, label
propagation~\cite{raghavan_near_2007} 
\end{itemize}
\fi

\subsection{Computation time management}
Three kinds of measures are computation-heavy in our experimental
setup : triangle computation, diameter and fb3. 
Fb3 needs $O(|C|^2)$ operations to compute the values for the community $C$.
$f_{clus}$ and $f_{perm}$ need the computation of all internal
triangles, which is very demanding for highly clustered graphs.

We therefore sample our dataset and average these two values over the sample.
We use the Hoeffeding bound~\cite{hoeffding_probability_1963} (our
samples are \textit{i.i.d} and in the $[0,1]$ interval) to get the
number of samples $t$ needed ensure that there is a small probability
$p$ that the error resulting in our sampling is not bounded by
$\epsilon$.
\begin{equation}
  P(|\overline{X} - E[X]| < \epsilon) = p \ge
  2e^{-2n\epsilon^2} \Leftrightarrow
  n \ge \frac{ln(p/2)}{-2\epsilon^2}
\end{equation}
We use 5000 samples, meaning that $p\le5\%$ and $\epsilon\le 0.02$.
Of course, the bound is a worst-case : in practice, we observe errors
of about $10^{-4}$, which is too small to disturb the rankings.

The diameter computation, needed by $f_{comp}$, is in $O(|C|^2)$.
We use the standard approximate algorithm based on two BFSs to compute
it in near-linear time.
The first BFS starts at a random point of the community, and the last
node visited by this BFS is used as the origin of another BFS. 
This heuristic searches for an eccentric point which is likely to
feature at the end of a maximum-distance path.

Due to the process of ranking quality functions and comparison
methods, even bounded errors may have unbounded impact on the results
if the approximated values are too close to each other.
On top of that, some of the community detection algorithms make
nondeterministic choices, which implies an incontrollable potential
difference in results.
To gain confidence that the randomness of the processes involved does
not influence the results too much, we ran the whole process multiple
times.
We obtained very close results in every run.

\section{Experimental results}

\subsection{Correlations in LFR}
We first study the results of the methodology when applied to LFR graphs.
In order to assess its stability, we create three benchmark graphs
from different random seeds for each class of LFR graphs described in
Sec.~\ref{sec:networks}.
In order to judge behavioral similarity of quality functions between
graphs, we compute Spearman's coefficient of each couple of graphs
(as presented in Sec.~\ref{sec:methodology},
step~\ref{item:spear-qual}) and report the results in
Tab.~\ref{table:NMI_LFR_GRAPHS}
(resp. Tab.~\ref{table:FB3_LFR_GRAPHS}) for NMI (resp. for fb3).
\noindent
\begin{table}[h]
\centering
\small
\caption{The correlation between the ranking of quality functions (with NMI ranking) for synthetic graphs\ifcolor\else\protect\footnotemark\fi}
\begin{tabular}{c|c|c|c|c|c|c|c|c|c|c|c|c|c|c|c}
file$\backslash$file &a1 &a2 &a3 &b1 &b2 &b3 &c1 &c2 &c3 &d1 &d2 &d3 &e1 &e2 &e3\\
a1 & - & \cc{1.00} & \cc{1.00} & \cc{0.98} & \cc{0.99} & \cc{0.45} & \cc{0.40} & \cc{-0.23} & \cc{-0.09} & \cc{0.53} & \cc{0.53} & \cc{0.53} & \cc{0.69} & \cc{0.31} & \cc{0.95}\\
a2 & - & - & \cc{1.00} & \cc{0.98} & \cc{0.99} & \cc{0.45} & \cc{0.40} & \cc{-0.23} & \cc{-0.09} & \cc{0.53} & \cc{0.53} & \cc{0.53} & \cc{0.69} & \cc{0.31} & \cc{0.95}\\
a3 & - & - & - & \cc{0.98} & \cc{0.99} & \cc{0.45} & \cc{0.43} & \cc{-0.22} & \cc{-0.07} & \cc{0.53} & \cc{0.53} & \cc{0.53} & \cc{0.69} & \cc{0.31} & \cc{0.94}\\
b1 & - & - & - & - & \cc{0.99} & \cc{0.48} & \cc{0.35} & \cc{-0.15} & \cc{0.01} & \cc{0.53} & \cc{0.53} & \cc{0.53} & \cc{0.69} & \cc{0.31} & \cc{0.92}\\
b2 & - & - & - & - & - & \cc{0.46} & \cc{0.34} & \cc{-0.21} & \cc{-0.07} & \cc{0.53} & \cc{0.53} & \cc{0.53} & \cc{0.69} & \cc{0.31} & \cc{0.94}\\
b3 & - & - & - & - & - & - & \cc{0.29} & \cc{0.03} & \cc{0.21} & \cc{0.15} & \cc{0.15} & \cc{0.15} & \cc{0.67} & \cc{0.56} & \cc{0.42}\\
c1 & - & - & - & - & - & - & - & \cc{0.07} & \cc{0.15} & \cc{0.27} & \cc{0.27} & \cc{0.27} & \cc{0.30} & \cc{0.25} & \cc{0.41}\\
c2 & - & - & - & - & - & - & - & - & \cc{0.90} & \cc{0.34} & \cc{0.34} & \cc{0.34} & \cc{-0.12} & \cc{0.05} & \cc{-0.32}\\
c3 & - & - & - & - & - & - & - & - & - & \cc{0.22} & \cc{0.22} & \cc{0.22} & \cc{-0.03} & \cc{0.20} & \cc{-0.16}\\
d1 & - & - & - & - & - & - & - & - & - & - & \cc{1.00} & \cc{1.00} & \cc{0.47} & \cc{0.12} & \cc{0.47}\\
d2 & - & - & - & - & - & - & - & - & - & - & - & \cc{1.00} & \cc{0.47} & \cc{0.12} & \cc{0.47}\\
d3 & - & - & - & - & - & - & - & - & - & - & - & - & \cc{0.47} & \cc{0.12} & \cc{0.47}\\
e1 & - & - & - & - & - & - & - & - & - & - & - & - & - & \cc{0.76} & \cc{0.70}\\
e2 & - & - & - & - & - & - & - & - & - & - & - & - & - & - & \cc{0.35}\\
e3 & - & - & - & - & - & - & - & - & - & - & - & - & - & - & -\\
\end{tabular}
\label{table:NMI_LFR_GRAPHS}
\end{table}
\ifcolor
\else
\footnotetext{A colored version is available on the authors' webpage.}
\fi

In Tab.~\ref{table:NMI_LFR_GRAPHS}, we see that quality functions of
the same class behave in the same way when compared to NMI. 
However, this positive view is tarnished by some exceptions : c1
seems to relate more to graphs of the \textit{a} class than from its
own class, and the same can be said from e3.
We assume that these exceptions are due to the random nature of the
generative model, which might produce networks that have some
structural properties that vary significantly enough to disturb
comparison with NMI.

It was expected that the \textit{a} class would be rated in the same
way as the \textit{b} and \textit{d} class, and would be different
from the other two. 
If the correct similarities are observed, surprisingly, the \textit{e}
class seems to behave similarly to the \textit{a} class, and this is
even clearer with the fb3 measure (cf
Tab.~\ref{table:FB3_LFR_GRAPHS}). 
This is probably because the distribution difference and the mixing
parameter have less influence in the structural properties of the
network than the overlapping nature.
We conclude that the comparison with NMI is globally efficient, but it
is very sensitive to noise and overlapping difference.
\noindent
\begin{table}[h]
\centering
\small
\caption{The correlation between the ranking of quality functions (with FB3 ranking) for synthetic graphs}
\begin{tabular}{c|c|c|c|c|c|c|c|c|c|c|c|c|c|c|c}
file$\backslash$file &a1 &a2 &a3 &b1 &b2 &b3 &c1 &c2 &c3 &d1 &d2 &d3 &e1 &e2 &e3\\
a1 & - & \cc{0.99} & \cc{1.00} & \cc{0.95} & \cc{0.94} & \cc{0.96} & \cc{-0.23} & \cc{-0.50} & \cc{-0.44} & \cc{0.72} & \cc{0.72} & \cc{0.72} & \cc{0.89} & \cc{0.89} & \cc{0.88}\\
a2 & - & - & \cc{0.99} & \cc{0.94} & \cc{0.93} & \cc{0.94} & \cc{-0.26} & \cc{-0.48} & \cc{-0.42} & \cc{0.70} & \cc{0.70} & \cc{0.70} & \cc{0.89} & \cc{0.87} & \cc{0.88}\\
a3 & - & - & - & \cc{0.95} & \cc{0.94} & \cc{0.96} & \cc{-0.23} & \cc{-0.50} & \cc{-0.44} & \cc{0.72} & \cc{0.72} & \cc{0.72} & \cc{0.89} & \cc{0.89} & \cc{0.88}\\
b1 & - & - & - & - & \cc{1.00} & \cc{1.00} & \cc{-0.26} & \cc{-0.49} & \cc{-0.44} & \cc{0.80} & \cc{0.80} & \cc{0.80} & \cc{0.90} & \cc{0.92} & \cc{0.93}\\
b2 & - & - & - & - & - & \cc{1.00} & \cc{-0.27} & \cc{-0.48} & \cc{-0.43} & \cc{0.80} & \cc{0.80} & \cc{0.80} & \cc{0.90} & \cc{0.92} & \cc{0.93}\\
b3 & - & - & - & - & - & - & \cc{-0.25} & \cc{-0.49} & \cc{-0.43} & \cc{0.80} & \cc{0.80} & \cc{0.80} & \cc{0.90} & \cc{0.91} & \cc{0.92}\\
c1 & - & - & - & - & - & - & - & \cc{0.62} & \cc{0.76} & \cc{-0.27} & \cc{-0.27} & \cc{-0.27} & \cc{-0.42} & \cc{-0.38} & \cc{-0.37}\\
c2 & - & - & - & - & - & - & - & - & \cc{0.90} & \cc{-0.26} & \cc{-0.26} & \cc{-0.26} & \cc{-0.49} & \cc{-0.51} & \cc{-0.37}\\
c3 & - & - & - & - & - & - & - & - & - & \cc{-0.31} & \cc{-0.31} & \cc{-0.31} & \cc{-0.47} & \cc{-0.47} & \cc{-0.39}\\
d1 & - & - & - & - & - & - & - & - & - & - & \cc{1.00} & \cc{1.00} & \cc{0.75} & \cc{0.78} & \cc{0.79}\\
d2 & - & - & - & - & - & - & - & - & - & - & - & \cc{1.00} & \cc{0.75} & \cc{0.78} & \cc{0.79}\\
d3 & - & - & - & - & - & - & - & - & - & - & - & - & \cc{0.75} & \cc{0.78} & \cc{0.79}\\
e1 & - & - & - & - & - & - & - & - & - & - & - & - & - & \cc{0.983} & \cc{0.963}\\
e2 & - & - & - & - & - & - & - & - & - & - & - & - & - & - & \cc{0.96}\\
e3 & - & - & - & - & - & - & - & - & - & - & - & - & - & - & -\\
\end{tabular}
\label{table:FB3_LFR_GRAPHS}
\end{table}

In Tab.~\ref{table:FB3_LFR_GRAPHS}, we see that the comparison with
fb3 is much more clear-cut than the one with NMI : there is no value
between -0.2 and 0.6, which would indicate medium to weak
correlations. 
It is also very clear that the \textit{c} class is considered as
differently ranked for fb3 than the other ones.
As stated above, the fb3 measure does not identify the model
difference in the generation of the \textit{e} class.

We note that the c1 and the e3 networks that did not behave like
the others when compared with NMI measure behave in the same way when
looking at the fb3 measure.
We conclude that comparing networks through the measure is less
sensitive than NMI to random variations due to network generation
processes, the downside being that it may show resemblance between two
networks that are actually very different.

\subsection{Correlations in real world data}
Just as with the LFR benchmark, we start by identifying groups of
networks where quality functions behave approximately in the same
way. 
Unlike LFR, the only classification available for these networks is
their representation of reality, and not the underlying model.

Real-life data are less clear-cut than controlled benchmark networks.
However, we see in Tab.~\ref{table:NMI_REAL_GRAPHS}
and~\ref{table:FB3_REAL_GRAPHS} that the connections (cora, CS) and
(lj, youtube, flickr) appear with both comparison methods as high,
which means that these networks are consistently close with the
ranking of their ground-truth.  This observation is consistent with
our knowledge of these networks.  Cora and CS both correspond to
scientific publication and their ground-truthes both correspond to
publication domains.  Interestingly, neither the overlapping nature of
CS nor the size difference seem to affect this outcome, which comforts
us in the robustness of the method.  Youtube, flickr and lj have
similar connection (someone follows someone) and ground-truth
(explicit membership) mechanics.  The other correlation relationships
differ given the considered comparison method.
\noindent
\begin{table}[h]
\centering
\caption{Spearman's coefficient of the rows of table~\ref{table:NMI_REAL} (NMI, Real-world)}
\begin{tabular}{c|c|c|c|c|c|c|c|c|c|c}
file$\backslash$file &CS &actors &amazon &cora &dblp &flickr &football &github &lj &youtube\\
CS & - & \cc{0.923} & \cc{0.281} & \cc{0.972} & \cc{0.302} & \cc{0.103} & \cc{0.245} & \cc{0.014} & \cc{0.253} & \cc{-0.187}\\
actors & - & - & \cc{0.264} & \cc{0.899} & \cc{0.276} & \cc{0.168} & \cc{0.318} & \cc{0.105} & \cc{0.262} & \cc{-0.077}\\
amazon & - & - & - & \cc{0.280} & \cc{0.965} & \cc{-0.231} & \cc{0.523} & \cc{-0.033} & \cc{-0.269} & \cc{-0.336}\\
cora & - & - & - & - & \cc{0.327} & \cc{0.115} & \cc{0.213} & \cc{0.052} & \cc{0.334} & \cc{-0.191}\\
dblp & - & - & - & - & - & \cc{-0.238} & \cc{0.453} & \cc{0.031} & \cc{-0.241} & \cc{-0.357}\\
flickr & - & - & - & - & - & - & \cc{0.180} & \cc{0.367} & \cc{0.808} & \cc{0.759}\\
football & - & - & - & - & - & - & - & \cc{0.350} & \cc{-0.191} & \cc{0.117}\\
github & - & - & - & - & - & - & - & - & \cc{0.329} & \cc{0.549}\\
lj & - & - & - & - & - & - & - & - & - & \cc{0.587}\\
youtube & - & - & - & - & - & - & - & - & - & -\\
\end{tabular}
\label{table:NMI_REAL_GRAPHS}
\end{table}
\paragraph{NMI :} We notice first that the tuple (cora, CS) is
extended to \textbf{(cora, CS, actors)}, which brings another
collaboration network close to the first two. 
We note, however, that the github network is not correlated with them.
We notice that the structural difference with github, where an
individual belongs to more groups than actors (7.8 compared to 3.8),
resembles the difference between LFR \textit{a} and \textit{c} class,
which was demoted by NMI. \\

An unexpected correlation is \textbf{(dblp, amazon)} : quality
functions behave in similar ways in a co-purchase network and in a
co-authorship network.  
As observed in Sec~\ref{sec:best-quality}, this result is due to the
erratic behavior of the correlation between qualities with very low
correlation values.
\noindent
\begin{table}[h]
\centering
\caption{Spearman's coefficient of the rows of table~\ref{table:FB3_REAL} (FB3, Real-world)}
\begin{tabular}{c|c|c|c|c|c|c|c|c|c|c}
file$\backslash$file &CS &actors &amazon &cora &dblp &flickr &football &github &lj &youtube\\
CS & - & \cc{-0.070} & \cc{0.920} & \cc{0.970} & \cc{0.502} & \cc{0.224} & \cc{-0.434} & \cc{-0.344} & \cc{-0.351} & \cc{-0.035}\\
actors & - & - & \cc{-0.052} & \cc{-0.157} & \cc{0.472} & \cc{-0.091} & \cc{0.776} & \cc{0.774} & \cc{0.434} & \cc{0.227}\\
amazon & - & - & - & \cc{0.935} & \cc{0.411} & \cc{0.189} & \cc{-0.455} & \cc{-0.378} & \cc{-0.316} & \cc{-0.105}\\
cora & - & - & - & - & \cc{0.409} & \cc{0.358} & \cc{-0.456} & \cc{-0.381} & \cc{-0.266} & \cc{0.040}\\
dblp & - & - & - & - & - & \cc{0.250} & \cc{0.163} & \cc{0.187} & \cc{0.143} & \cc{0.456}\\
flickr & - & - & - & - & - & - & \cc{0.154} & \cc{0.156} & \cc{0.533} & \cc{0.790}\\
football & - & - & - & - & - & - & - & \cc{0.911} & \cc{0.719} & \cc{0.497}\\
github & - & - & - & - & - & - & - & - & \cc{0.654} & \cc{0.414}\\
lj & - & - & - & - & - & - & - & - & - & \cc{0.760}\\
youtube & - & - & - & - & - & - & - & - & - & -\\
\end{tabular}
\label{table:FB3_REAL_GRAPHS}
\end{table}
\paragraph{FB3 :} We notice a surprising correlation of the
co-purchase network with scientific networks \textbf{(amazon, cora,
CS)}.

The networks that are strongly defined by the underlying bipartite
network, \textbf{(football, actor, github)}, are correlated with fb3. 
They have a similar structure, with a particularly high clustering
coefficient inside of the communities. 

We observe that the \textbf{(lj, football, github)} tuple appears as
close to each other.  It could be explained by the underlying
bipartite model of lj (and the other two OSNs) that creates a weak
correlation with the other networks that are structurally more defined
by it.


\subsection{Quality functions in contexts}
\label{sec:best-quality}
We analyze the correlations between the quality functions and the
comparison methods. 
Our aim is to find quality functions that give a consistent ranking
that is highly correlated with the ground truth.

\noindent
\begin{table}[h]
\centering
\caption{Spearman's coefficient of the NMI(ground truth, algorithms) compared to the results of quality functions. Real-world dataset}
\begin{tabular}{c|c|c|c|c|c|c|c|c|c|c|c|c}
file$\backslash$quality &cc &fb3 &mod &nmi &perm &sign &cond &FOMD &comp &cut\_ratio &f-odf &sur\\
CS & \cc{0.00} & \cc{0.82} & \cc{-0.25} & \cc{1.00} & \cc{0.00} & \cc{-0.14} & \cc{0.14} & \cc{0.61} & \cc{-0.04} & \cc{0.32} & \cc{0.00} & \cc{-0.46}\\
actors & \cc{-0.54} & \cc{0.46} & \cc{-0.89} & \cc{1.00} & \cc{-0.21} & \cc{-0.50} & \cc{-0.21} & \cc{-0.21} & \cc{-0.39} & \cc{-0.21} & \cc{-0.32} & \cc{-0.57}\\
amazon & \cc{-0.30} & \cc{0.03} & \cc{-0.97} & \cc{1.00} & \cc{-0.97} & \cc{-0.12} & \cc{-0.97} & \cc{-0.87} & \cc{0.03} & \cc{-0.96} & \cc{-0.97} & \cc{-0.44}\\
cora & \cc{0.06} & \cc{0.69} & \cc{-0.06} & \cc{1.00} & \cc{0.06} & \cc{-0.06} & \cc{0.19} & \cc{0.69} & \cc{0.06} & \cc{0.44} & \cc{0.19} & \cc{-0.06}\\
dblp & \cc{-0.43} & \cc{0.89} & \cc{-0.96} & \cc{1.00} & \cc{-0.96} & \cc{-0.32} & \cc{-0.89} & \cc{-0.57} & \cc{0.18} & \cc{-0.88} & \cc{-0.86} & \cc{-0.46}\\
flickr & \cc{0.00} & \cc{-0.71} & \cc{0.75} & \cc{1.00} & \cc{0.61} & \cc{0.07} & \cc{0.61} & \cc{0.07} & \cc{0.14} & \cc{-0.01} & \cc{0.61} & \cc{-0.75}\\
football & \cc{0.38} & \cc{0.38} & \cc{0.10} & \cc{1.00} & \cc{0.88} & \cc{0.38} & \cc{-0.37} & \cc{0.56} & \cc{0.38} & \cc{-0.87} & \cc{-0.33} & \cc{0.38}\\
github & \cc{-0.29} & \cc{-0.36} & \cc{-0.11} & \cc{1.00} & \cc{-0.07} & \cc{0.11} & \cc{-0.11} & \cc{-0.04} & \cc{-0.43} & \cc{-0.14} & \cc{0.07} & \cc{-0.04}\\
lj & \cc{-0.21} & \cc{-0.86} & \cc{0.43} & \cc{1.00} & \cc{0.21} & \cc{-0.18} & \cc{0.50} & \cc{0.25} & \cc{0.32} & \cc{0.35} & \cc{0.50} & \cc{-0.32}\\
youtube & \cc{0.36} & \cc{-0.89} & \cc{0.96} & \cc{1.00} & \cc{0.79} & \cc{0.39} & \cc{0.68} & \cc{0.07} & \cc{0.11} & \cc{0.31} & \cc{0.68} & \cc{0.61}\\
\end{tabular}
\label{table:NMI_REAL}
\end{table}

\noindent
\begin{table}[h]
\centering
\caption{Spearman's coefficient of the fb3(ground truth, algorithms) compared to the results of quality functions. Real-world dataset}
\begin{tabular}{c|c|c|c|c|c|c|c|c|c|c|c|c}
file$\backslash$quality &cc &fb3 &mod &nmi &perm &sign &cond &FOMD &comp &cut\_ratio &f-odf &sur\\
CS & \cc{-0.50} & \cc{1.00} & \cc{-0.14} & \cc{0.82} & \cc{0.14} & \cc{-0.75} & \cc{0.39} & \cc{0.75} & \cc{-0.61} & \cc{0.59} & \cc{0.18} & \cc{-0.93}\\
actors & \cc{0.29} & \cc{1.00} & \cc{-0.07} & \cc{0.46} & \cc{-0.79} & \cc{0.43} & \cc{-0.79} & \cc{-0.79} & \cc{0.18} & \cc{-0.79} & \cc{-0.64} & \cc{0.36}\\
amazon & \cc{-0.86} & \cc{1.00} & \cc{-0.04} & \cc{0.03} & \cc{-0.04} & \cc{-0.89} & \cc{0.00} & \cc{0.25} & \cc{-0.93} & \cc{-0.01} & \cc{0.00} & \cc{-0.79}\\
cora & \cc{-0.64} & \cc{1.00} & \cc{0.04} & \cc{0.69} & \cc{0.29} & \cc{-0.75} & \cc{0.50} & \cc{0.89} & \cc{-0.75} & \cc{0.79} & \cc{0.50} & \cc{-0.75}\\
dblp & \cc{-0.68} & \cc{1.00} & \cc{-0.79} & \cc{0.89} & \cc{-0.79} & \cc{-0.57} & \cc{-0.64} & \cc{-0.32} & \cc{-0.07} & \cc{-0.67} & \cc{-0.61} & \cc{-0.71}\\
flickr & \cc{0.18} & \cc{1.00} & \cc{-0.21} & \cc{-0.71} & \cc{0.07} & \cc{0.04} & \cc{0.07} & \cc{0.46} & \cc{0.39} & \cc{0.60} & \cc{0.07} & \cc{0.29}\\
football & \cc{1.00} & \cc{1.00} & \cc{0.68} & \cc{0.38} & \cc{0.25} & \cc{1.00} & \cc{-0.96} & \cc{-0.05} & \cc{1.00} & \cc{-0.21} & \cc{-0.93} & \cc{1.00}\\
github & \cc{-0.29} & \cc{1.00} & \cc{0.39} & \cc{-0.36} & \cc{-0.57} & \cc{0.71} & \cc{-0.61} & \cc{-0.79} & \cc{0.71} & \cc{-0.57} & \cc{-0.93} & \cc{0.71}\\
lj & \cc{0.29} & \cc{1.00} & \cc{-0.54} & \cc{-0.86} & \cc{-0.14} & \cc{0.39} & \cc{-0.46} & \cc{-0.36} & \cc{-0.11} & \cc{-0.38} & \cc{-0.46} & \cc{0.32}\\
youtube & \cc{0.04} & \cc{1.00} & \cc{-0.86} & \cc{-0.89} & \cc{-0.61} & \cc{-0.07} & \cc{-0.54} & \cc{0.04} & \cc{0.14} & \cc{-0.19} & \cc{-0.54} & \cc{-0.32}\\
\end{tabular}
\label{table:FB3_REAL}
\end{table}

In the context of OSNs (flickr, youtube, lj), we see in
Tab.~\ref{table:FB3_REAL} that the fb3 does not give us an answer on
the best quality function to use since no satisfying correlation is
observed. 
However, we see in Tab.~\ref{table:NMI_REAL} that NMI tells us that
Modularity gives a consistently correlated score, while Permanence
also behaves well while being more inconsistent (notably with lj). 

Concerning scientific collaboration networks (cora, CS), the average
FOMD consistently shows a strong correlation with the ground-truth,
close to the Cut-ratio. 
This tendency is coherent with both comparison methods.

The networks with strong bipartite underlying structure (football,
github, actor) do not show any particular outlier when compared with
NMI with very weak correlations. 
However, fb3 outlines the performance of Signature and Surprise.

The last two networks, amazon and dblp, do not show any satisfying
correlation with the selected quality functions. We suspect the
quality functions that we use are not adapted to the \texttt{contexts}
of these graphs.

\section{Conclusion}
In this paper, we introduced fb3 as a clustering comparison method for
community detection algorithms.
We gave evidence that quality functions
are \texttt{context}-dependant.
The application of a quality function comparison methodology resulted
in the identification of three \texttt{contexts} and of the relevant
quality functions. 
We also provided evidence that the methodology clearly differentiate
\texttt{contexts}.

The methodology that has been presented here may very well be applied
to overlapping/weighted quality functions that would measure the
efficiency of overlapping/weighted community detection algorithms. 

We are currently in the process of integrating all the functionalities
presented in this paper in a tool that will be made
available shortly to the public.

\bibliographystyle{splncs}
\bibliography{biblio}

\end{document}